\documentclass[reprint,
superscriptaddress,
showkeys,
 amsmath,amssymb,
 aps,
]{revtex4-2}

\usepackage{graphicx}
\usepackage{dcolumn}
\usepackage{bm}
\usepackage{xcolor}
\usepackage{xr}
\begin{document}

\preprint{APS/123-QED}

\title{Macroscopic Dominance from Microscopic Extremes: Symmetry Breaking in Spatial Competition}

\author{Stuti Guha}
\email {sg22dr@fsu.edu}
\affiliation{Department of Mathematics, Florida State University, Tallahassee, FL, 32306}
\author{Shawn D. Ryan}%
 \email{s.d.ryan@csuohio.edu}
\affiliation{Department of Mathematics and Statistics, Cleveland State University, Cleveland, OH, 44115
}%
\affiliation{Center for Applied Data Analysis and Modeling, Cleveland State University, Cleveland, OH, 44115}


\author{Bhargav R. Karamched}
\email{bkaramched@fsu.edu}
\affiliation{Department of Mathematics, Florida State University, Tallahassee, FL, 32306}%
\affiliation{Institute of Molecular Biophysics, Florida State University, Tallahassee, FL, 32306
}%
\affiliation{Program in Neuroscience, Florida State University, Tallahassee, FL, 32306}



\begin{abstract}
How do competing populations convert a spatial advantage into macroscopic dominance? We introduce a stochastic model for resource competition that decouples the transient discovery phase from monopolization. Initial symmetry breaking is governed by extreme value statistics of first-passage times: a linear spatial disadvantage requires an exponentially larger population to overcome. However, transient superiority cannot stabilize dominance. A non-reciprocal interaction bias is strictly necessary to arrest local fluctuations and drive the system into a robust absorbing state.
\end{abstract}

\maketitle


\noindent \textit{Introduction.} 
Competition for limited resources is a fundamental driver of spatial organization and survival in ecosystems~\cite{lehman1997competition,yodzis2013competition}. When multiple distinct populations vie for a localized resource, the dynamics frequently drive the system toward mutual exclusion, culminating in a macroscopic absorbing state where a single group monopolizes the target~\cite{Bai23,zhang2021winner,simberloff1982status}. Classical ecological models often rely on deterministic mass-action kinetics to predict these outcomes~\cite{van2010deciphering,swailem2023lotka,bianchi2017landscape,dercole2016ecology}, but such frameworks fail to capture the inherent stochasticity of spatial exploration~\cite{tauber2025stochastic}. In reality, the initial symmetry breaking in spatial competition is fundamentally a first-passage time problem, driven entirely by microscopic extreme events. This leads to a universal scaling law: a colony with a linear spatial disadvantage requires an exponentially larger population to compensate.

\begin{figure*}
    \centering
    \includegraphics[width = \textwidth]{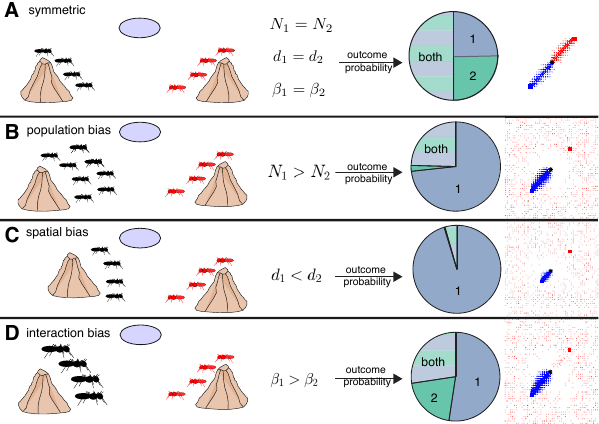}
    \caption{Competition outcomes under symmetry breaking. Pie charts show the final-state probability of the absorbing outcome: colony 1 dominance, colony 2 dominance, or coexistence (“both”). (A) Symmetric conditions ($N_1 = N_2$, $d_1 = d_2$, $\beta_1 = \beta_2$) produce a nonuniform distribution of outcomes; (B–D) Breaking symmetry through (B) population bias, (C) spatial bias, and (D) interaction bias shifts the outcome probabilities. Outcomes summarize 1000 trials. Rightmost column shows representative lattice equilibria in each case.}
    \label{fig:figure1}
\end{figure*}

To investigate the dynamics of this competitive exclusion, we introduce and analyze a hybrid model of active matter and collective behavior, where stochastic agent-based motion is coupled to a reaction-diffusion process.  We specifically frame this model in the context of discrete ant colonies foraging for a common food source. Ant foraging represents a classic paradigm of distributed biological problem-solving: individual agents effectively perform unbiased random walks until a resource is located, at which point they deposit a chemical pheromone trail. The diffusion and evaporation of this pheromone field create a dynamic chemotactic gradient, triggering a feedback loop that rapidly transitions the colony from disordered exploration to highly directed, collective exploitation. There is also an additional interesting dynamic due to competition between colonies for scarce resources once the colony reaches a mature state (over 1-2 years) \cite{gordon1995development}.  Members of each colony have the ability to secrete unique pheromones so that nest mates can distinguish friend from foe  based on complex hydrocarbons \cite{martin2008colony,guerrieri2009ants,brandt2009scent,van2010deciphering,torres2007role}.

By systematically perturbing this system, we investigate how macroscopic dominance is dictated by three orthogonal asymmetries: spatial proximity, population size, and interaction bias (``fear'' or avoidance of a competitor's chemical trail). We demonstrate that the initial symmetry breaking is governed by the extreme first-passage statistics of the earliest foragers, establishing a dimensionless scaling law that dictates the topological boundaries of the competition. However, we show that this fluctuation-driven discovery phase is insufficient to guarantee victory. Transient spatial or numerical superiority cannot stabilize dominance; rather, nonlinear chemical feedback is necessary to arrest local fluctuations and drive the system into a robust absorbing state. By combining extreme first passage statistics with nonlinear interaction bias, our model captures a universal phenomenon observed across ecological conflicts: a profound spatial disadvantage or a strong psychological aversion can be overcome by an exponential advantage in numbers~\cite{jarymowicz2006dominance,tedeschi2021fear}.\\




\noindent \textit{Model Description.} We build on the model we developed in Ref.~\cite{hartman2024walk}. We model the general terrain as an $M \times M$ lattice and track the dynamics of $\mathcal{N} \in \mathbb{N}$ distinct ant colonies. For $1 \leq i \leq \mathcal{N}$, we denote by $N_i$ the population size of colony $i$. We assume that food discovery and trail formation occur on a sufficiently fast time scale and ignore births and deaths in the colonies. There is no direct communication between individuals, but rather a response to a chemical pheromone gradient if present represents the only means of (indirect) communication.  We designate a single site as the nest, $\mathbf{x}^i_0$, for each ant colony. Initially, all $N_i$ ants of each colony occupy their nest. To understand what factors affect competition for a single food source between distinct colonies, we also randomly designate a single site $\mathbf{x}_f$ as a food source. We describe the distance of a colony's nest to the food source with the parameter $d_i \equiv ||\mathbf{x}_i^0 - \mathbf{x}_f||$. 

Each colony of ants is represented by a set of points $\{\mathbf{x}^{(i)}_j(t)\}$, $j = 1,..., N_i$, where $\mathbf{x}^{(i)}_j(t) \in [1,M] \times [1,M]$ and $t$ is time. Each point can be thought of as the location of the center of mass for an individual ant. We assume the boundaries are reflecting for ants to maintain a fixed population.

Ant motion is subdivided between two types of ants: (i) foraging ants and (ii) carrying ants. Foraging ants are those that are searching for a food source. They undergo a random walk modeled as a Brownian motion \cite{Pop23,Cha21}. That is, a foraging ant at a given site moves to any adjacent site in its Moore neighborhood with equal probability.  
At each time step an ant picks a new site to move to based on this probability and thus all ants move with a constant speed. A foraging ant becomes a carrying ant once it locates a food source. We assume that ants that reach a food source begin carrying food. Their dynamics are no longer Brownian.  Instead, such carrying ants make a direct path for the nest \cite{Mul88,Weh03,Weh16}.  Upon reaching the nest, a carrying ant again becomes a foraging ant.

To recruit additional ants to the food source a given ant found, a carrying ant secretes pheromone {in exponentially decreasing amounts as it travels to its nest to form a chemical trail from the nest to the food source. We assume ants keep secreting pheromone until they reach the nest in contrast to another recent work that has considered ants having a finite chemical supply until they ``recharge" at the nest site \cite{Mal15}.  The physical response of an ant to a chemical stimulus is well documented \cite{Cal92b,CouFra03,panait2004pheromone}.  Crucially, foraging ants in our lattice model incorporate logarithmic gradient sensing, mirroring Weber’s Law of perception \cite{weber1972gardening,cammaerts2020weber,Amo19,perna2012individual}. This ensures that the competitive dynamics are driven not by absolute chemical concentrations, but by the relative strengths of the competing pheromone fields, a pervasive feature of biological information processing.  The rate of pheromone secretion is assumed to diffuse and decrease exponentially with a carrying ant's distance from the food location~\cite{CouFra03,robinson2008decay}. Pheromone deposition and trail laying are modeled by a two-dimensional reaction-diffusion process for the chemical concentration corresponding to colony $i$, $c_i(\mathbf{x},t)$, first introduced in \cite{Rya16} and later adapted to multiple colonies in \cite{RyaBau20}:
\begin{equation}
\frac{\partial c_i}{\partial t} - {D}{\Delta_d c_i} + \gamma c_i  = \sum_{j=1}^{J_i} {Ae^{-{||\mathbf{x}^{(i)}_j(t) - \mathbf{x}_f||}^2}}{\delta {(\mathbf{x} - \mathbf{x}^{(i)}_j(t))}}.
\label{eq1}
\end{equation}

Here $J_i$ is the number of carrying ants from colony $i$, $\Delta_d$ is the discrete Laplacian, $D$ is the diffusion coefficient controlling the rate at which the pheromone spreads, and $\gamma$ is the evaporation coefficient that ensures an exponential decay of the pheromone in time. The coefficient $A e^{-||\mathbf{x}^{(i)}_j(t) - \mathbf{x}_f||^2}$ represents the amount of pheromone deposited at time $t$ and decays as a carrying ant moves away from the food source. This decrease is needed to ensure that the proper gradient forms due to the competition with diffusion. The initial distribution of chemical is taken as zero so there is no pre-defined directional preference. We prescribe homogeneous Fourier-type boundary conditions for the pheromone so that we have
\begin{align*}
-D\frac{\partial c_i}{\partial x} \Big|_{x=0} &= - c_i(0,y,t) \quad
-D\frac{\partial c_i}{\partial x} \Big|_{x=M} = c_i(M,y,t)\\
-D\frac{\partial c_i}{\partial y} \Big|_{y=0} &= - c_i(x,0,t)\quad
-D\frac{\partial c_i}{\partial y} \Big|_{y=N} =  c_i(x,N,t).
\end{align*}
Hence, Fickian flux along the boundary is preserved. That is, along the boundary, the chemical continues to flow with its concentration gradient.\\

\noindent \textit{Pheromone sensing.} In the absence of any pheromone, a foraging ant moves into any of the sites in its Moore neighborhood with equal probability. 

Foraging ants subject to the pheromone concentration field undergo a biased Brownian motion. The probability that a foraging ant of colony $i$ moves to a specific adjacent site in its Moore neighborhood increases if pheromone produced by colony $i$ is greater at that site relative to the pheromone concentration at the current location of the ant. Conversely, if the pheromone produced by a colony $j \neq i$ is greater at that site relative to the pheromone concentration at the current location of the ant, the probability the ant moves to that site decreases. {This is consistent with recent studies showing indirect decision making by individual ants in response to local pheromone gradients (seen both in experiments of \cite{CouFra03,mokhtari2022spontaneous} and theoretical studies \cite{panait2004pheromone}).}  To implement this, we compute the difference in pheromone concentration at the current location of the ant, $\mathbf{x}^c$, and the pheromone concentration at a given adjacent site in the Moore neighborhood of the ant, $\mathbf{x}^a$. Here, $a \in A$ {and $c \notin A$}, where $A$ is the set of locations in the Moore neighborhood of the ant selected to move. {Thus, $\mathbf{x}_c \neq \mathbf{x}_a$}. Let $\Delta c^a_i \equiv c_i(\mathbf{x}^a,t) - c_i(\mathbf{x}^c,t)$. We assign a weight, $\mathcal{W}_i^a$, to each site as follows:
$$
\mathcal{W}_i^a =
\frac{1 + \Delta c^a_i}{1 + \sum_{k\neq i}^{\mathcal{N}}{\beta^i_k \Delta c^a_k}},
$$
where $\beta^i_k$ is a parameter representing how much colony $i$ fears colony $k$.
The probability that a foraging ant of colony $i$ subject to a pheromone concentration field moves from $\mathbf{x}_c$ to $\mathbf{x}_a$ is then
$$
\mathcal{P}_{\mathbf{x}_c \to \mathbf{x}_a}^i = \frac{\mathcal{W}^a_i}{\sum_{k \in A} \mathcal{W}^k_i}.
$$

For details on simulating the above system, please refer to the Supplementary Material (SM)\footnote{The Supplementary Material (SM) contains details of the simulation of the model and calculation of the scaling laws.}. With the model established, we seek to identify the primary factors dictating macroscopic competitive fitness.\\

\noindent \textit{Two colony competition.} To answer this question, we first consider the case where $\mathcal{N} = 2$, which is analytically tractable and yields insights into the general case. We characterize a colony's dominance of a food source as an outcome where only that colony forms a trail to the food source and the other does not. We do not consider simultaneous trail formation by distinct colonies as a successful outcome. Such situations lead to ant battles, which drain resources to obtain the food~\cite{champer2024battles,adler2018mechanistic}. This yields four possible outcomes for each simulation: (1) colony 1 dominates, (2) colony 2 dominates, (3) both form trails, or (4) neither form trails.

Two colonies having equal opportunity to win a food source yields a nonuniform distribution of outcomes(Fig.~\ref{fig:figure1}A):  stochastic fluctuations select among degenerate absorbing states. Predominantly, both colonies will form simultaneous trails. We investigate how three symmetry-breaking orthogonal factors modify the equal opportunity outcome probabilities: population size, proximity to food, and fear factor ($\beta$). Biasing any of these three factors modifies the outcome probability in a different way (Fig.~\ref{fig:figure1}B-D), indicating that collective selection is governed by multiple mechanisms rather than a single effective fitness.

The proximity of a colony to the food source most strongly increases colony fitness in acquiring food, followed by population size, and fear factor last (see Fig.~\ref{fig:figure1}). This hierarchy manifests because proximity most strongly impacts the time for the first ant of a given colony to locate the food source, followed by the population size. Fear only comes into play after pheromone secretion begins, so it does not affect the time for the first ant to find the food source.

\begin{figure*}
    \centering
    \includegraphics{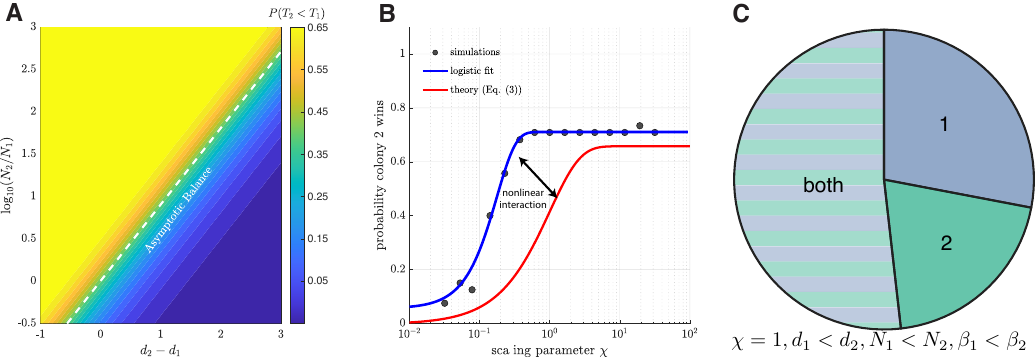}
    \caption{The scaling parameter $\chi$ completely determines symmetry. (A) Phase diagram of $\mathbb{P}(T_2 < T_1)$ based on the full lattice simulations. The dashed white line denotes the asymptotic balance condition ($\chi = 1$), where an exponential population advantage perfectly cancels a linear spatial disadvantage. Each point in the heatmap emerged from averaging over 10000 trials. (B) Probability that colony 2 dominates the food source. (C) Pie chart demonstrating the reemergence of symmetry with $\chi = 1$ in the absorbing outcomes despite unequal parameters.  Simulation parameters: lattice size $M = 50 \times 50$, $D = 10$, $\gamma = 2$, $N_1 8^{-d_1} = 0.43$.}  
    \label{fig:figure2}
\end{figure*}

The hierarchy of these symmetry-breaking mechanisms is fundamentally governed by the extreme statistics of the initial search process. To see this, let $T_i$, $i = 1,2$, represent the extreme first passage time for an ant from colony $i$ to discover the food source. Further assume that $d_1 < d_2$ and $N_1 < N_2$. Because the model random walks are discrete, $T_i \geq d_i$, and in particular we have~\cite{karamched2026entropic} 
\begin{equation}
    \langle T_i\rangle = d_i + (1-p_i)^{N_i} +o(1),
    \label{eq:extreme}
\end{equation}
where $p_i$ represents the probability that an ant takes the shortest Moore path to the food source and $p_i = M_i8^{-d_i}$. The entropic parameter $M_i$ quantifies the number of shortest paths connecting the nest to the food source.

We approximate the probability that $T_2 < T_1$ as $\mathbb{P}(T_2<T_1) \approx \mathbb{P}(T_2 = d_2)\mathbb{P}(T_1>d_1) = \left[1-(1-p_2)^{N_2}\right](1-p_1)^{N_1}$.
For this probability to remain non-negligible, we require an asymptotic balance between the colonies:
$$
(1-p_1)^{N_1} \sim (1-p_2)^{N_2}
$$
For sufficiently large $\min\{N_1, N_2\}$, and assuming $p_i \ll 1$, we have that 
\begin{equation}
  \frac{N_2}{N_1} \sim \frac{M_1}{M_2}z^{d_2-d_1},
  \label{eq:scaling}
\end{equation}
where $z$ is the effective geometric coordination number. For Moore neighborhoods, $z = 8$. For Von Neumann neighborhoods, $z = 4$ (see SM).

Equation~\eqref{eq:scaling} says that a colony with a linear disadvantage in proximity requires an exponentially greater number of ants to compensate. Thus, being closer to food is `exponentially better' than having more ants.

It also suggests that the key parameter underlying symmetry in the system is the dimensionless scale parameter
\begin{equation}
    \chi \equiv \frac{N_2}{N_1}8^{d_1-d_2},
    \label{eq:key_parameter}
\end{equation}
which quantitates the advantage of a colony: $\chi > 1$ means colony 2 is at an advantage and $\chi < 1$ means colony 1 is at an advantage. The entropic parameters satisfy $M_1 \sim M_2$ because the space is homogeneous, and thus factor out of $\chi$. Indeed, the curve parameterized by $\chi = 1$ in parameter space bifurcates the heatmap describing $\mathbb{P}(T_2<T_1)$ into regions where colony 2 finds the food first or second (Fig. ~\ref{fig:figure2}A.)

While our extreme-value theory yields the exact dimensionless scaling parameter governing time to locate the food source, the strict asymptotic probability limits underestimate the simulated outcomes for macroscopic food domination. This deviation is predominantly due to the nonlinear interactions between the colonies driven by pheromone secretion and the resulting fear factor. However, plotting the simulated probabilities against $\chi$ reveals an empirical data collapse (Fig.~\ref{fig:figure2}B), demonstrating that $\chi$ fundamentally governs the macroscopic competition, even in the presence of lattice artifacts and nonlinear interactions. Indeed, setting $\chi = 1$ with unequal colony parameters yields a reemergence of symmetry (Fig.~\ref{fig:figure2}C).

\begin{figure*}
    \centering
    \includegraphics[width = \textwidth]{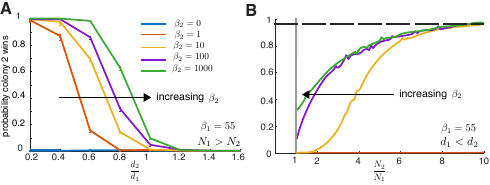}
    \caption{Asymmetric interaction bias controls the sharpness of the competitive phase transition. (A) The probability of Colony 2 establishing macroscopic dominance as a function of the relative distance ratio $d_2/d_1$, given a strict numerical disadvantage ($N_1 > N_2$) and a fixed competitor bias ($\beta_1 = 55$). (B) The probability of dominance as a function of the population ratio $N_2/N_1$, given a strict spatial disadvantage ($d_1 < d_2$) and constant $\beta_1 = 55$. As indicated by the arrows, increasing Colony 2's interaction bias ($\beta_2$) shifts the competitive outcome from a noise-dominated, probabilistic regime into a strictly polarized state. For large $\beta_2$, the symmetry-breaking boundary steepens into a sharp phase transition, enabling Colony 2 to overcome structural deficits to secure the resource and drive the system into the absorbing state of competitive exclusion.}
\label{fig:figure3}
\end{figure*}

Discrete spatial constraints also impose a rigid physical limit on competitive advantage. While the scaling parameter $\chi$ suggests that an exponentially large population can overcome any spatial deficit, the discrete lattice enforces an absolute ``speed limit'' on discovery. Because an ant cannot traverse the grid faster than one site per time step, the closer resident colony (Colony 1) possesses a guaranteed minimum discovery time of $d_1$ steps. Consequently, even as the competing population diverges ($\chi \to \infty$), their probability of dominance does not asymptote to unity. Instead, it is strictly bounded by the stochastic failure rate of the more proximal colony, yielding a theoretical ceiling: $\lim_{\chi \to \infty} \mathbb{P}(T_2 < T_1) = \exp(-N_1 p_1)$ (see SM). This reveals a fundamental ecological constraint: infinite numerical superiority cannot overcome the absolute physical limits of a highly optimized resident. A massive invading force can only capitalize on the resident's stochastic errors.


\textit{The Significance of $\beta$.} The interaction bias, $\beta$, plays a paradoxical but decisive role in the symmetry breaking of the system. Because $\beta$ is a strictly secondary, post-discovery mechanism, it is entirely inactive during the extreme first-passage time bottleneck of the initial foragers. However, while it cannot alter the topological scaling limit of discovery, it strictly governs the stability of the subsequent macroscopic state. 

Figure~\ref{fig:figure3} demonstrates that $\beta$ acts as a tuning parameter for the sharpness of the competitive phase transition. In the absence of a strong interaction bias ($\beta \to 0$), the competitive outcome remains a noise-dominated, probabilistic regime. In this state, the system is highly vulnerable to steady-state flux; a colony cannot stabilize its hold on the resource and requires an overwhelming proximity advantage to maintain dominance. 

Conversely, increasing the asymmetric interaction bias steeply polarizes the symmetry-breaking boundary. As seen in Figure~\ref{fig:figure3}A, for sufficiently large $\beta$, the probability profile sharpens into a step-like function. This non-reciprocal interaction violently arrests local fluctuations, allowing a colony to entirely override an opponent's initial first-discovery advantage. Figure~\ref{fig:figure3}B further illustrates that this interaction-driven phase transition can secure the absorbing state even against a competitor with a strict initial spatial advantage. Ultimately, this demonstrates an important sociological and ecological universality: a profound structural or spatial deficit can be completely overcome by a sheer, exponential advantage in numbers, provided the non-linear interaction barrier is sufficiently strong~\cite{jarymowicz2006dominance,tedeschi2021fear}.\\

\noindent \textit{Conclusion.} Our model for resource competition puts forth three orthogonal factors that break the symmetry of competition outcome with distinct mechanisms. Extreme value statistics of the first passage time to find the resource suggest that proximity and population size predominantly govern the initial symmetry breaking and discovery of resource, with proximity to the resource imparting the greatest influence. Our analysis indicates that a linear disadvantage in proximity for a colony of ants may be overcome with exponentially more ants than the competitor. The interaction bias, or fear factor in this Letter, does not affect a colony's ability to find a resource, imputing the smallest symmetry-breaking effect upon absorbing states. On the other hand, it repels competitor ants from the food source, and in its absence a colony has no chance to dominate the resource. Thus, it is a necessary condition for resource dominance.

The interaction bias imputed by chemical reaction-diffusion dynamics are not unique to ants. Bacteria and slime molds also exhibit emergent behavior stemming from, for example, quorum sensing~\cite{striednig2022bacterial,popat2015collective,vijayreddy2024single,reid2016collective}. Our flexible model can therefore be modified to describe a large class of systems whose global interaction occurs via chemical sensing. 

While this model explicitly analyzes pairwise competition, the underlying mechanics naturally extend to $\mathcal{N}$-colony ecosystems due to a strict separation of timescales. Because the discovery phase is governed by extreme value statistics, the probability of simultaneous first arrivals among $\mathcal{N}$ colonies is vanishingly small. Consequently, the first colony to arrive establishes the initial interaction bias, effectively transforming an $\mathcal{N}$-body spatial race into a sequence of pairwise exclusionary events. The repulsive chemical barrier established by the macroscopic victor renders the remaining $\mathcal{N}-1$ competitors dynamically equivalent to a single, aggregate suppressed state.

In an ecological context, our framework mathematically formalizes the distinction between discovery and exploitation. Recent work also shows that coexistence is a viable state when foraging strategies are increasingly different \cite{perfecto1994foraging}.  Winning the race to a resource is purely a stochastic optimization problem governed by proximity; however, monopolizing that resource is fundamentally a game of spatial exclusion. A population cannot simply out-search a competitor to secure dominance. Without the repulsive force of the interaction bias to structurally sever the competitor's access, even the fastest and most populous colony remains forever vulnerable to competitive noise.

\end{document}